\documentclass[aps,prl,twocolumn,,groupedaddress,superscriptaddress,showpacs]{revtex4}
\usepackage{amssymb}
\usepackage{amsmath}
\usepackage{graphicx}
\usepackage{epstopdf}

\input{tcilatex}

\begin{document}

\title{ High-Symmetry Polarization Domains in Low-Symmetry Ferroelectrics}
\author{I. Lukyanchuk}
\affiliation{LPMC, University of Picardie Amiens, France and Landau Institute for
Theoretical Physics, Russia}
\author{P. Sharma}
\affiliation{Department of Physics and Astronomy, Nebraska Center for Materials and
Nanoscience, University of Nebraska, Lincoln, NE 68588, USA}
\author{T. Nakajima}
\affiliation{Institute of Materials Research, Tohoku University, Sendai, 980-8577 Japan}
\author{S. Okamura}
\affiliation{Department of Applied Physics, Tokyo University of Science, Tokyo, 162-8601,
Japan}
\author{J. F. Scott}
\affiliation{Cavendish Laboratory, 19 JJ Thomson Ave, Cambridge, UK}
\author{A. Gruverman}
\affiliation{Department of Physics and Astronomy, Nebraska Center for Materials and
Nanoscience, University of Nebraska, Lincoln, NE 68588, USA}
\date{\today }

\begin{abstract}
We present experimental evidence for hexagonal domain faceting in the
ferroelectric polymer PVDF-TrFE films having the lower orthorhombic
crystallographic symmetry. This effect can arise from purely electrostatic
depolarizing forces. We show that in contrast to magnetic bubble shape
domains where such type of deformation instability has a predominantly
elliptical character, the emergence of more symmetrical circular harmonics
is favored in ferroelectrics with high dielectric constant.
\end{abstract}

\pacs{77.80.Dj, 77.55.fp, 68.37.Ps} \maketitle

In nature, low-symmetry molecules often self-assemble to produce
high-symmetry mesoscopic arrays or bundles of colloids, nanoparticles,
proteins, and viruses \cite{Glotzer2012}. Similar kind of mesoscopic
self-assembly occurs for domain pattern formation in ferroelectric crystals.
However, in most cases the symmetry of such domains is either the same or
lower than the crystal symmetry and is fixed in space by the
crystallographic axes.

There are, however, very few reports where domain faceting has a
greater symmetry than that of the lattice or even incompatible with
it. Triangular and
hexagonal domains have been observed in lead germanate Pb$_{5}$Ge$_{3}$O$%
_{11}$ \cite{Shur1989} and lithium niobate LiNbO$_{3}$
\cite{Scrymgeour2005} crystals with 3-fold polar axes. Relaxation of
initially circular domains into pentagonal and hexagonal ones was
also reported in [111]-oriented lead zirconate-titanate
Pb(Zr,Ti)O$_{3}$ films \cite{Ganpule2002}. In some cases these
effects  occur under  external perturbation, such as laser or e-beam
irradiation. A 6-fold pattern in laser-irradiated ferroelectric of PbMg$%
_{1/3}$Nb$_{2/3} $O$_{3}$, which has a purely cubic symmetry, was
observed by Scott \emph{et al.} \cite{Scott1996a,Scott1996b}. In
earlier studies, Schwarz and Hora reported the $6$-fold diffraction
patterns appearing along a $2$-fold axis in quartz irradiated
simultaneously by e-beams and blue laser light
\cite{Schwarz1969,Schwarz1970}. Brown and Hollingsworth reported
domain structures in thiourea compounds with  $12$-fold symmetry in
spite of the fact that the crystals had only $2$-fold lattice symmetry \cite%
{Brown1995}. A peculiar kind of hexagonal domain crystallization has been
observed in sodium dodecyl sulfate surfactant \cite{Flesselles1991},
but there is only a rather general understanding of these phenomena.

\begin{figure}[b]
\centering
\includegraphics [width=6.2 cm] {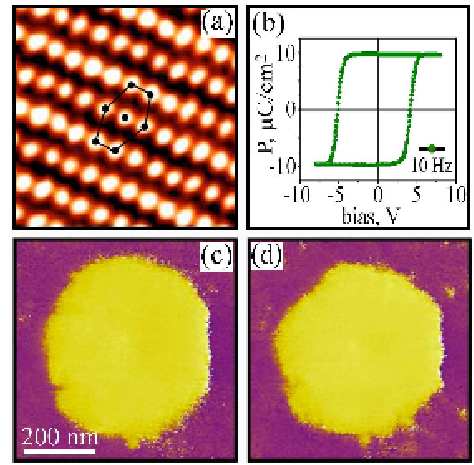} \hspace{0 cm}
\caption{(a) An atomically-resolved STM image of the surface of PVDF-TrFE
film at room temperature (Reprinted with permission from \protect\cite%
{Cai2004}. (b) Polarization hysteresis loop of PVDF-TrFE film. (c) PFM image
of the domain written by a -55 V, 1s pulse applied by a PFM tip. (d) PFM
image of the same domain 15 minutes after pulse application.}
\label{FigExp}
\end{figure}

In this work, we present new data on hexagonal domain formation in the
ferroelectric co-polymer of poly(vinylidene fluoride-trifluoroethylene
(PVDF-TrFE) that lacks any hexagonal lattice symmetry. We interpret our data
in terms of the electrostatic depolarization instability-theory developed by
Thiele \cite{Thiele1969} to depict the elliptical deformation of small
circular domain patterns in ferromagnetic bubble-memory devices \cite%
{Malozemoff1979, Keller1987}.\ However, unlike magnetic bubbles,
ferroelectric PVDF-TrFE favors the hexagonal faceting instability.

PVDF-TrFE is the best known ferroelectric polymer widely used due to
its relatively high polarization and piezoelectric parameters
\cite{Lovinger1983}. Discovery of ferroelectricity in ultrathin
PVDF-TrFE films \cite{Ducharme2000} and fabrication of highly
ordered nanomesas arrays for non-volatile memories \cite{Hu2009}
draw additional attention due to possibility of integration of these
materials into all-organic electronic systems. It remains a model
material for investigation of the mechanism of polarization reversal
in organic ferroelectrics.

The PVDF-TrFE (75/25) copolymer films have been fabricated by spin-coating
the VDF/TrFE solution dissolved in diethyl carbonate (3 wt\%) on the (111)
Pt-sputtered ($150\,\mathrm{nm}$) on Ti($5\,\mathrm{nm}$)/SiO$_{2}$($200\,%
\mathrm{nm}$)/Si substrate followed by annealing at $120\,^{\circ }\mathrm{C}
$. The thickness of the PVDF-TrFE film is $50\,\mathrm{nm}$. X-ray $\theta $-%
$2\theta $ scans showed a clear peak at $19.7^{\circ }$, which corresponded
to (110) $d$-spacing of the all-trans ferroelectric $\beta $-phase \cite%
{Bai2006}. The (110) peak  pointed out to the highly textured PVDF-TrFE
molecular chains lying parallel to the substrate.

A commercial atomic force microscope (Asylum MFP-3D) has been used to
generate and visualize the domain structure in the piezoresponse force
microscopy (PFM) mode. Domain visualization has been performed by applying
1.5 V modulating voltage in the $200-400\,\mathrm{kHz}$ frequency range to
the Pt-Ti-coated silicon cantilevers. We used a ferroelectric test system
(Radiant system) to record the $P-E$ hysteresis loops from the Au top
electrodes deposited by thermal evaporation.

Ferroelectric polarization in PVDF-TrFE arises from alignment of molecular
dipoles formed by electropositive hydrogen and electronegative fluorine
atoms. The all-trans ($\beta $-phase) molecular conformation results in the
dipoles being oriented nearly perpendicular to the chain axis and
polarization reversal is associated with rotation of these dipoles about the
molecular chains. In the crystalline $\beta $-phase of highly textured
PVDF-TrFE, molecules tend to pack parallel to each other forming a structure
with the orthorhombic symmetry $Cm2m$ ($C_{2v}^{14}$) space group, with
2-fold polar axis nearly perpendicular to the film's plane. In Fig.\ref%
{FigExp}a, it can be seen that the molecular chains arranged themselves in
parallel rows, forming a quasihexagonal close packing structure \cite%
{Cai2004}.

Preliminary PFM testing showed that the as-grown PVDF-TrFE films were
uniformly polarized downward. Polarization loops were shifted horizontally
toward the negative voltage (Fig.\ref{FigExp}b) suggesting a presence of an
internal built-in electric field oriented toward the substrate. This field
could be estimated from the polarization loop asymmetry as $E=(\left\vert
V_{c}^{+}\right\vert -\left\vert V_{c}^{-}\right\vert )/\varepsilon h \simeq
10^{5}\,\mathrm{V/cm}$, where $V_{c}^{-}$, and $V_{c}^{+}$ are negative and
positive coercive voltages, respectively, $h\simeq 50$ $\mathrm{nm}$ is film
thickness and $\varepsilon \simeq 15$ is the dielectric permittivity \cite%
{Lovinger1983}. Figures \ref{FigExp}c,d illustrate the domain faceting
effect observed in these films. A voltage pulse in the range from $-15\,%
\mathrm{V}$ to $-60\,\mathrm{V}$ applied to the PFM tip results in a single
circular domain with radius $r$ of about $50-250$ $\,\mathrm{nm}$ (Fig.\ref%
{FigExp}c) with polarization oriented against the imprint. After the field
is off, the domain starts to slowly relax back acquiring a hexagonal shape
during the process (Fig.\ref{FigExp}d).

We show that both the domain contraction and tendency to expand its
perimeter via faceting instability are provided by the purely electrostatic
force balance.

It is known that the depolarizing effect of surface charges due to
discontinuity of polarization vector at the interface plays the critical
role in the formation and dynamics of ferroelectric domains. However, the
foreseeable periodic Landau-Kittel domain structure \cite%
{Landau1935,Kittel1946} with expected half-period $d\simeq \left( h\xi
_{0}\right) ^{1/2}\simeq 10$ $\mathrm{nm}$ (where $\xi _{0}\simeq 2$ $%
\mathrm{nm}$ is the coherence length) was not observed in these PVDF-TrFE
films. Most probably, this is due to the fact that the depolarization field
is initially screened by accumulation of the charged species at the top
surface and by the charge carriers of the conductive electrode at the bottom
of the film.

During application of the switching pulse by the PFM tip, the surface
screening charges in the vicinity of the tip are dispersed, activating the
depolarization forces in the area of the newly-generated domain. Subjected
to electrostatic tension, which could also stem from the internal built-in
field, this domain starts to relax until it is stabilized by the
redistributed surface screening charges.

To account for the forces, acting on the generated domain just after its
creation and to reveal the \ faceting instability, we, following \cite%
{Thiele1969}, distinguish three contributions to the energy $W_{r}$ of
cylindrical domain:%
\begin{equation}
W_{r}=W_{dw}+W_{PE}+W_{d}  \label{TotEnergy}
\end{equation}

(i) $W_{dw}$ is the domain wall (DW) energy, which is provided by DW surface
tension $\sigma _{dw}=\varepsilon ^{-1}\left( 4\pi P^{2}\right) \xi _{0}$:
\begin{equation}
W_{dw}\simeq \varepsilon ^{-1}\left( 4\pi P^{2}\right) \xi _{0}\left( 2\pi
r\right) h.  \label{Wdw}
\end{equation}%
However, the $W_{dw}$  is at least of $\xi _{0}/h \simeq 25$ times
smaller than two other contributions (unlike in magnetic domains
where $\xi _{0}\sim 50nm$) and can be neglected in electrostatic
calculations.

(ii) $W_{PE}$ is the energy of polarization interaction with the electrical
field $E$, which in our case of the relaxing domain is the internal built-in
imprint field:
\begin{equation}
W_{PE}=2\left( \pi r^{2}h\right) PE.  \label{WPE}
\end{equation}%
This contribution is proportional to the domain volume $\pi r^{2}h$ and
factor $2$ is introduced because of the domain re-polarization.

(iii) $W_{d}$ is the depolarization energy, which is determined by
the surface charge of density $\sigma =\pm P$ due to polarization
termination at the film surface. To outline how this contribution
drives the domain shape instability we note that $W_{d}$ is nothing
else, but the energy excess of the cylindrical capacitor with the
plate charges $Q=\pm\left( 2P\right) \pi r^{2} $ which is oppositely
polarized with respect to the original
ferroelectric slab. Capacitance of such finite-size capacitor with $%
\varepsilon =1$ was first calculated by Kirchhoff in 1877 \ \cite%
{Kirchhoff1877} in the limiting case of $r\gg h$ and two years later
generalized by Lorentz\ \cite{Lorentz1879} in terms of the elliptic function
for arbitrary aspect ratio $h/r$. In late of 1960's, Thiele performed
similar calculations for the energy of the cylindrical magnetic domain \cite%
{Thiele1969}, but again with $\varepsilon =1$ \cite{Bratkovsky2000a}. We
summarize their results for the case of $r\gg h$ as:%
\begin{equation}
W_{d1}\simeq -\left( 2\pi r\right) h^{2}\left( 2P\right) ^{2}\ln \frac{8r}{%
e^{1/2}h},  \label{Kirchhoff}
\end{equation}%
where $e\simeq 2.71$ and the index "one" means $\varepsilon =1$ \cite%
{Bratkovsky2000a}.

Expression (\ref{Kirchhoff}) is similar to the well known energy of
the edge fringing field of the finite-size capacitor
\cite{Kirchhoff1877,Lorentz1879}. Importantly, the negative sign of
$W_{d1}$ is explained by the repulsion of the edge-forming dipoles.
Caused by the long-range electrostatic forces, this term is not
simply proportional to the DW surface, like  but also contains the
non-local logarithmic correction, dependent on the integral domain
shape.  Acting against
 $W_{PE}$ which minimizes the domain volume, it determines the tendency of
DW to increase its surface by either faceting or roughening
deformations.

The size and shape of domain is, therefore, the result of a balance between
contributions of different terms in Eq.(\ref{TotEnergy}). The comprehensive
analysis of possible domain instabilities has been carried out by Thiele for
ferromagnetic materials, which are characterized by the almost uniform
permittivity \cite{Thiele1969}. To apply these results to ferroelectric
films, one has to find out how Thiele's approach can be generalized for the
case of the films with arbitrary dielectric constant $\varepsilon >1$.

Let us consider two dipoles with charges $\pm q$, located across the slab of
permittivity $\varepsilon $ and thickness $h$ and separated by a distance $%
\rho $. Their interaction energy can be calculated as \cite{Supplimentary}:
\begin{equation}
W_{d\varepsilon }=-\frac{4\varepsilon ^{2}}{\varepsilon ^{2}-1}%
\sum_{n=1}^{\infty }\left( -\beta \right) ^{n}W_{1d}(nh),\quad \beta =\frac{%
\varepsilon -1}{\varepsilon +1},  \label{Wimmage}
\end{equation}%
where $W_{d1}(h)=2q^{2}\left[ r^{-1}-\left( h^{2}+\rho ^{2}\right) ^{-1/2}%
\right] $ is the energy of dipolar interaction in vacuum at
$\varepsilon =1$. Since each domain can be considered as an ensemble
of interacting parallel dipoles, Eq.(\ref{Wimmage}) can also be used
to find the depolarization
energy of such a domain $W_{d\varepsilon }$ for an arbitrary value of $%
\varepsilon $ if its energy $W_{d1}$ is known for $\varepsilon =1$.

Using summations $\sum_{n=1}^{\infty }n^{2}(-\beta )^{n}=-\frac{\varepsilon
^{2}-1}{4\varepsilon ^{3}}$ and $c_{\varepsilon }=\frac{4\varepsilon ^{2}}{%
\varepsilon ^{2}-1}\sum_{n=1}^{\infty }\left( -\beta \right) ^{n}n^{2}\ln n%
\overset{{\varepsilon \gg 1}}{\simeq }{0.84}$, we transform Eq.(\ref%
{Kirchhoff}) to:%
\begin{equation}
W_{d\varepsilon }\simeq -\left( 2\pi r\right) h^{2}\left[ \frac{1}{%
\varepsilon ^{2}}\ln \frac{8r}{e^{1/2}h}+\frac{c_{\varepsilon }}{\varepsilon
}\right] \left( 2P\right) ^{2}\text{.}  \label{Wdeps}
\end{equation}

Transformation (\ref{Wimmage}) is valid for a single domain in a
free-standing ferroelectric film. To account for the influence of conducting
substrate, one should apply the mirror boundary condition, provided by one
more transformation $W_{d}(h)\rightarrow \frac{1}{2}W_{d}(2h)$ that we shall
use at the end.

The equilibrium domain radius is given by the minimum of the energy $%
W_{PE}+W_{d\varepsilon }$. A closer look at Eqs (\ref{WPE}) and (\ref{Wdeps}%
) reveals that this minimum lies well below the actual radius, at $r\ll h$
and is comparable to the Kittel domain width $d\simeq 10$ $\mathrm{nm}$.
This explains the tendency of the newly created domain to relax via viscous
contraction.

To take into consideration the possibility of shape-breaking instability
during relaxation process we parameterize the domain boundary in polar
coordinates as $R=R(\theta )$ and, following \cite{Thiele1969}, account its
deviation from the cylinder with $R=r$ by emergence of circular harmonics: \
\begin{equation}
R(\theta )=r+\sum_{k=2}^{\infty }\left( \Delta r_{k}\right) \cos k\theta ,
\label{harm}
\end{equation}%
with small amplitudes $\left\vert \Delta r_{k}\right\vert \ll r$.

To address the question of domain stability under perturbation (\ref{harm})
we, following \cite{Thiele1969}, expand the variation of the energy as:
\begin{equation}
W=W_{r}+\frac{1}{2}\sum_{k=2}^{\infty }\left[ W_{PE}^{\prime \prime
}+W_{d}^{\prime \prime }(k)\right] \left( \Delta r_{k}\right) ^{2},
\label{Wstab}
\end{equation}%
where
\begin{equation}
W_{PE}^{\prime \prime }=\frac{1}{2}\frac{\partial ^{2}W_{PE}}{\partial r^{2}}%
=2\pi hPE  \label{WPEpp}
\end{equation}%
is the harmonics interaction with imprint field and $W_{d}^{\prime \prime
}(k)$ is the depolarization field contribution. For $\varepsilon =1$ and in
the limit $r\gg h$ this term was calculated in \cite{Thiele1969} as:
\begin{equation}
W_{d1}^{\prime \prime }(k)\simeq -\pi \left( 2P\right) ^{2}\frac{h^{2}}{r}%
k^{2}\ln \frac{8er}{kh},  \label{bk1}
\end{equation}%
Here, we neglected the DW energy contribution as mentioned above.

A cylindrical domain is unstable with respect to axial symmetry-breaking
modes if the corresponding coefficients in (\ref{Wstab}) are negative:
\begin{equation}
W_{PE}^{\prime \prime }+W_{d}^{\prime \prime }(k)<0.  \label{ineqv}
\end{equation}%
The instability comes from the negative logarithmic term in
(\ref{bk1}). However to explore the instability condition
(\ref{ineqv}) we should generalize the expression (\ref{bk1}) for
arbitrary $\varepsilon >1$ using the transformation (\ref{Wimmage}).
Summing the log-terms in lowest in $h/r$ order we obtain:
\begin{equation}
W_{d\varepsilon }^{\prime \prime }(k)=-\frac{\pi h^{2}}{r}\left[ \frac{1}{%
\varepsilon ^{2}}\ln \frac{8er}{kh}+\frac{c_{\varepsilon }}{\varepsilon }%
\right] \left( 2P\right) ^{2}k^{2}.  \label{bkeps}
\end{equation}%
Now using the transformation: $W_{d\varepsilon }^{\prime \prime
}(k,h)\rightarrow \frac{1}{2}W_{d\varepsilon }^{\prime \prime }(k,2h)$ we
include the effect of the bottom electrode and arrive at the instability
condition:

\begin{equation}
\frac{\varepsilon E}{P}\frac{r}{(2h)}<2k^{2}\left[ \frac{1}{\varepsilon }\ln
\frac{8er}{k(2h)}+c_{\varepsilon }\right] ,  \label{GraphEq}
\end{equation}%
For $\varepsilon \gg 1$, we can neglect the log term and reduce condition (%
\ref{GraphEq}) to the simpler form: $\frac{\varepsilon E}{P}\frac{r}{h}%
<3.4k^{2}$.

\begin{figure}[t]
\centering
\includegraphics [width=7 cm] {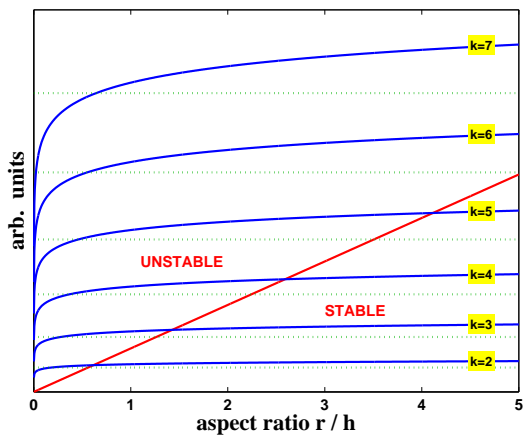} \hspace{0 cm}
\caption{Graphical solution of instability condition (\protect\ref{GraphEq})
for PVDF film with $\protect\varepsilon \simeq 15$ \protect\cite%
{Lovinger1983}, $\protect\varepsilon E/P\simeq 1.6$ as function of the
aspect ratio $r/h$. Dotted lines present the approximate solutions for high $%
\protect\varepsilon $.}
\label{FigInstab}
\end{figure}

Graphic analysis shown in Fig.\ref{FigInstab} reveals that the
shape-breaking instability develops first for the high $k$-modes,
which have a \textit{higher} symmetry than the crystal background.
Notably, for a domain with initial radius $r\simeq 4h\simeq
200\,\mathrm{nm}$, the
hexagonal faceting instability with $k=6$ can occur, whereas the modes with $%
k<4$ and, in particularly, elliptic deformation with $k=2$,
compatible with two-fold symmetry of PVDF-TrFE  are
electrostatically protected. \ The situation is quite different from
the magnetic case were the elliptical instability is the most
favorable one \cite{Malozemoff1979,Thiele1969}. The reason is that,
the domain formation in ferroelectrics is stabilized by the
bulk field-interaction term (\ref{WPE}) whereas the DW surface-tension term (%
\ref{Wdw}) appears to be more relevant in ferromagnets.

However, to know which of the electrostatically allowed symmetry-breaking
modes is unstable during domain relaxation we should go beyond the
linear-stability analysis, since several factors are coming into the play.
For instance, the DW pinning in the atomic planes shown in Fig.\ref{FigExp}a
can be critical for the mode selection. Minimization of the anisotropic
pinning energy, know as Wulff construction \cite{Wulff1901}, acting along
with the high-$k$ electrostatic instability, can favor selection of those
surfaces that are most close to the perfect hexagonal faceting.

The dynamic buckling effects during domain contraction can also favor the
particular (hexagonal) shape-breaking as was shown, for example, for the
case of the relaxing drops in dipolar ferrofluids \cite{Langer1992}.

To conclude, we have discovered hexagonal faceting of domains in
ferroelectric polymer PVDF-TrFE films, which have symmetry much lower than
hexagonal. We interpret this effect in terms of the electrostatic
depolarization instability originally used by Thiele for magnetic bubble
domains. Unlike magnetic bubbles with the $k=2$ elliptical instability,
ferroelectrics favor the instability with higher-$k$ modes. Our analysis
raises questions about the origin of high-symmetry domain faceting in Pb$%
_{5} $Ge$_{3}$O$_{11}$ \cite{Shur1989} and PZT \cite{Ganpule2002}
ferroelectrics. The described approach can be applied to a wider range of
instabilities in quite unrelated materials, including $6$-fold faceting in
surfactants \cite{Flesselles1991} and in thiourea inclusion compounds with $%
k=12$ \cite{Brown1995}.

\newpage

\emph{SUPPLEMENTARY MATERIAL}

 \textbf{Dipolar Interaction via Dielectric
Slab}

\begin{figure}[h]
\centering
\includegraphics [width=7 cm] {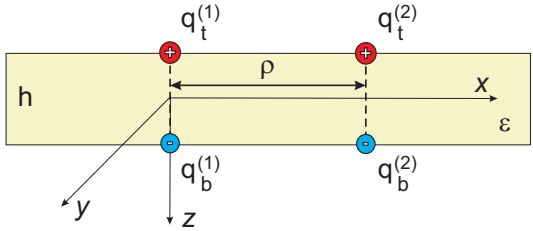} \hspace{0 cm}
\caption{Dipolar location geometry and coordinate system}
\label{FigSupl}
\end{figure}

We calculate the interaction energy of two dipoles, located across
the dielectric slab of thickness $h$ and permittivity $\varepsilon $
at distance $\rho $ from each other, as shown in Fig.~\ref{FigSupl}.
The coordinate
system ($x,y,z$) is placed in the center of the first dipole, axis $\mathbf{z%
}$ being directed downward, perpendicular to the slab and axis $\mathbf{x}$%
\textbf{\ }being in the direction of the second dipole. The
coordinates of the top charge $q_{t}^{(1)}=+q$ and of the bottom
charge  $q_{b}^{(1)}=-q$
of the first dipole \ are $\mathbf{R}_{t}^{(1)}=(0,0,-h/2)$ and $\mathbf{R}%
_{b}^{(1)}=(0,0,h/2)$ correspondingly. The corresponding coordinates
of the
top and bottom charges of the second dipole, $q_{t}^{(2)}=+q$  and $%
q_{b}^{(2)}=-q$, are $\mathbf{R}_{t}^{(2)}=(\rho ,0,-h/2)$ and $\mathbf{R}%
_{b}^{(2)}=(\rho ,0,h/2)$.

Let us consider  the field due to the upper positive charge of the
first dipole $q_{t}^{(1)}=+q$. The distribution of the electrostatic
potential inside the slab can be presented by the integral relation,
involving the zero-order Bessel function $J_{0}(r_{\perp })$
\cite{Batygin1978} :

\begin{equation}
\varphi _{t}^{(1)}=\int_{0}^{\infty }\left[ B_{1}(k)e^{-k\left( z+\frac{h}{2}%
\right) }+B_{2}(k)e^{k\left( z+\frac{h}{2}\right) }\right]
J_{0}(kr_{\perp })dk,  \label{PhiQ}
\end{equation}%
with $r_{\perp }=(x^{2}+y^{2})^{1/2}$, $-\frac{h}{2}<z<\frac{h}{2}$,
\begin{equation}
B_{1}=q\frac{\left( 1-\beta \right) }{1-\beta ^{2}e^{-2kh}},\quad B_{2}=q%
\frac{\beta \left( 1-\beta \right) e^{-2kh}}{1-\beta ^{2}e^{-2kh}}
\end{equation}%
and
\begin{equation}
\beta =\frac{\varepsilon -1}{\varepsilon +1}.
\end{equation}

Substitution of $B_{1}$ and $B_{2}$ into (\ref{PhiQ}) gives:%
\begin{equation}
\varphi _{t}^{(1)}=q\left( 1-\beta \right) \int_{0}^{\infty }\frac{%
e^{-k\left( z+\frac{h}{2}\right) }+\beta e^{k\left( z-\frac{3h}{2}\right) }}{%
1-\beta ^{2}e^{-2kh}}J_{0}(kr_{\perp })dk.
\end{equation}

Similarly, the distribution of potential of the bottom negative
charge of the first dipole $q_{b}^{(1)}=-q$ can be written as
$\varphi _{b}^{(1)}(r_{\perp },z)=-\varphi _{t}^{(1)}(r_{\perp
},-z)$ and the resulting potential distribution of the first dipole
$\varphi _{d}^{(1)}=\varphi _{t}^{(1)}+\varphi _{b}^{(1)}$ as:
\begin{equation}
\varphi _{d}^{(1)}=q\left( 1-\beta \right) \int_{0}^{\infty }\frac{%
e^{-kz}-e^{kz}}{1+\beta e^{-kh}}e^{-kh/2}J_{0}(kr_{\perp })dk.
\end{equation}

The electrostatic energy of interaction between first and second
dipole is
calculated as:%
\begin{gather}
W_{d\varepsilon }=\varphi _{d}^{(1)}(\mathbf{R}_{t}^{(2)})\,q_{t}^{(2)}+%
\varphi _{d}^{(1)}(\mathbf{R}_{b}^{(2)})\,q_{b}^{(2)}  \label{En} \\
=2q^{2}\left( 1-\beta \right) \int_{0}^{\infty
}\frac{1-e^{-kh}}{1+\beta e^{-kh}}J_{0}(k\rho )dk.  \notag
\end{gather}

To simplify Eq.(\ref{En}) we expand the denominator in series:%
\begin{equation}
\frac{1}{1+\beta e^{-kh}}=\sum_{n=0}^{\infty }\left( -\beta \right)
^{n}e^{-khn}
\end{equation}%
and integrate the resulting expression, using the formula:%
\begin{equation}
\int_{0}^{\infty }e^{-k\lambda }J_{0}(k\eta )dk=\frac{1}{(\eta
^{2}+\lambda ^{2})^{1/2}},\quad \lambda >0.
\end{equation}%
After some algebra, taking into account that $\sum_{n=1}^{\infty
}\left( -\beta \right) ^{n}=-\left( 1+\beta ^{-1}\right) ^{-1}$, we
obtain
\begin{gather}
W_{d\varepsilon }(h)=2q^{2}\left( 1-\beta \right) \left[ \frac{1}{r}%
+\sum_{n=1}^{\infty }\frac{\left( 1+\beta ^{-1}\right) \left( -\beta
\right)
^{n}}{(\rho ^{2}+n^{2}h^{2})^{1/2}}\right]  \\
=-\frac{4\varepsilon }{\varepsilon ^{2}-1}\sum_{n=1}^{\infty }\left(
-\beta \right) ^{n}W_{d1}(nh),  \notag
\end{gather}%
where
\begin{equation}
W_{d1}(h)=2\left[ \frac{q^{2}}{\rho }-\frac{q^{2}}{(\rho ^{2}+h^{2})^{1/2}}%
\right]
\end{equation}%
is the energy of dipolar interaction in vacuum at $\varepsilon =1$.

\end{document}